%% file: Gravitational_Radiation_and_Charges_on_de_Sitter_proceedings_2.tex
\documentclass[a4paper]{jpconf}
\usepackage{graphicx}
\usepackage{iopams}

\usepackage{amsmath}
\usepackage{amssymb}
\usepackage{amsfonts}
\usepackage{wrapfig}

\input{globaldef}

\newcommand{\scri}{\mathcal{I}}

\newcommand{\dQadd}{{\dot{Q}^{\mbox{\textit{add}}}}_D}

\newcommand{\Qadd}{{Q^{\mbox{\textit{add}}}}}

\begin{document}
\title{Gravitational Radiation and Charges on de Sitter}

\author{Emine Şeyma Kutluk}

\address{Scuola Normale Superiore, Pisa, Italy}

\ead{emine.kutluk@sns.it}

\begin{abstract}
We give a summary of our work \cite{dS-rad-1} and \cite{dS-rad-2} with Geoffrey Compère and Sk.Jahanur Hoque. After quickly reviewing the $\Lambda$-BMS solution space and why it is not trivial to define an analogue of Bondi mass loss, we discuss how this can be resolved by studying the linearized solutions. Then we describe how to solve the linearized Einstein's equations in the generalized harmonic gauge and the set of assumptions such as adiabaticity and being at a large distance from the source that allows us to write the metric perturbations in terms of the source multipole moments. We point out how to consistently truncate the multipole expansion. We then discuss the net change of the metric perturbation at future infinity due to a source that is time-varying in a finite time interval that will be connected to the memory effect. We explain how to define charge flows using the ``holographic stress-tensor" and what form they take in terms of multipole moments at second order in perturbations. We conclude with a some remarks on the results.

\end{abstract}

%

\section{Introduction}

First proposed in early 1900s, gravitational radiation/waves and their effects opens up one of the most important window to the gravitational physics.
One of the best-known description of them is via the Bondi-Metzner-Sachs (BMS) formalism \cite{bms}, which \textit{fully non-linearly} (but with a Taylor approximation) describes the solutions to Einstein's vacuum equations with boundary conditions that allow for describing a gravitational radiation.
This analysis was generalized recently to allow for a cosmological constant \cite{Compere-LambdaBMS}.
However in this analysis it has not been possible to define charges that is similar to the Bondi mass.
In this note we will discuss the papers \cite{dS-rad-1} and \cite{dS-rad-2} where one considers linearized gravitational radiation around the de Sitter (dS) background, express it in terms of the multipole moments of the source that are \textit{consistently truncated}, and propose charges that includes a ``proper" mass.
We will also discuss a displacement memory effect in terms of the source multipole moments.\\

\section{Asymptotically de Sitter Solutions}

Let us start by a quick review of the ($\Lambda$)-BMS procedure/solutions \cite{Compere-LambdaBMS}. The upshot is that the metric 
\begin{equation}
ds^2=e^{2\beta} \frac{V}{r}du^2-2e^{2\beta}dudr+g_{AB}(dx^A-U^Adu)(dx^B-U^Bdu)
\end{equation} 
on a four dimensional Lorentzian manifold where $A,B=1,2$ numbers the angular coordinates and where
\begin{align}
& g_{AB}=r^2 q_{AB}(u,x^C)+ r \ C_{AB}(u,x^C) {+\frac{1}{4} q_{AB}(u,x^C) C^2} + \frac{1}{r} E_{AB}(u,x^C) {+ \mathcal{O} ( r^{-2} ) \ , }\\
& \frac{V}{r} = {\frac{\Lambda}{3}r^2 - \frac{R[q]}{2} + \frac{\Lambda}{16} C_{AB}(u,x^C)C^{AB}(u,x^C) - \frac{2}{r} {M(u,x^C)} + o(r^{-1})  \ , }\\
&\beta ={ - \frac{1}{32r^2} C^{AB}(u,x^C) C_{AB}(u,x^C) + \mathcal{O} ( r^{-4} ) \ , }\\
&U^A = { - \frac{1}{2r^2} D_B C^{AB}(u,x^C) - \frac{2}{3r^3} \lp {N^A}(u,x^C) - \frac{1}{2} C^{AB}(u,x^C) D^C C_{CB}(u,x^C) \rp + \mathcal{O} ( r^{-4} ) \ }  , 
 \end{align}
\footnote{Here $o(r^{-1})$ means decays faster than $r^{-1}$, in contrast to $\mathcal{O}(r^{-1})$ which would mean decays as $r^{-1}$ or faster.}solves the source-free Einstein's equations with cosmological constant
\begin{equation}
G_{\mu \nu}+\Lambda g_{\mu \nu}=0
\end{equation}
\footnote{Gravitational constant is taken to be 1.}if they satisfy the conditions we will describe in a second. $D$ is the $q_{AB}$ covariant derivative, and raising and lowering is also performed by $q_{AB}$ and its inverse. $C^2:= C^{AB}C_{AB}$ and $R[q]$ is the Ricci scalar of the leading order angular metric $q_{AB}$, whereas $C_{AB}$ and $E_{AB}$ are subleading corrections. The fields $N^A(u,x^C)$ and $M(u,x^C)$ arise as integration constants.  We have made an $r^{-1}$ expansion to capture effects from an isolated source.  Moreover we have set some of the coefficients to zero by a proper gauge choice.(See \cite{Compere-LambdaBMS} for more details.)\\
To solve the Einstein equations $C_{AB}$ needs to satisfy
\begin{align}
& \frac{\Lambda}{3} C_{AB}= \pr_u q_{AB} \ 
\end{align} 
whereas $M$ and $N_A$ need to satisfy the \textit{flux-balance laws}
\begin{align}
& \partial_u N_A^{(\Lambda)} - \partial_A M^{(\Lambda)} + \frac{\Lambda}{2}D^BJ_{AB}=0 \ ,\\
& \partial_u  M^{(\Lambda)}+ \frac{\Lambda}{6} D^A N_A^{(\Lambda)}+ \frac{\Lambda^2}{24} C^{AB} J_{AB}=0 \ ,
\end{align} 
where
\begin{align}\label{eq:M-Lambda}
& M^{(\Lambda)}=M +  \partial_u C_{CD} C^{CD} + \cdots \ , \\
& N_A^{(\Lambda)}=N_A {-\frac{3}{2\Lambda}D^B\partial_u C_{AB} + \cdots \ ,} \\
& J_{AB}=-E_{AB}{-\frac{2}{\Lambda^2} \partial^2_u C_{AB} + \cdots \ .}
\end{align} 
(See \cite{Compere-LambdaBMS} for full definitions.) Let us look at the spatial integral of the $\partial_u M$ equation
\begin{equation}
\partial_u \int_{S^2} M=- \frac{1}{8} \int \partial_uC_{AB} \partial_u C^{AB} + \frac{\Lambda^2}{24} \int C^{AB} E_{AB} + \Lambda [\cdots] \ .
\end{equation} 
For the flat case $\Lambda=0$, a strict mass loss can be seen  but the same is not true for $\Lambda \neq 0$ case, due to the existence of \textit{free} $E_{AB}$ term. In the next section we will see that linearizing gravity and connecting the metric to the source moments by linearized Einstein equations $E_{AB}$ will no longer be independent, allowing us to have a proper mass definition.

\section{Linearized Solution and Multipole Expansion}
With the realization of previous section, we now take up the mission of discussing the asymptotically dS spacetime in the linearized approximation. For this we will change gears and solve in the first subsection the linearized Einstein's equations in a new gauge, now with a source. As we will see the equations get a very simple form in this gauge. In this analysis we will put in the source since want to connect the metric components to a radiating source. We will solve for Einstein's equation using proper Green's function and use further assumptions to simplify our expressions.
In the second subsection we define multipole moments of the source and discuss how to consistently truncate the multipole expansion. We finally write down the metric in terms of the first few multipole moments and discuss the effect of the radiating source in the far future.

\subsection{Linearized Solution and the Approximations}
Let us now consider the linearization, and further connect the metric components to a matter source. Starting with the Einstein equation
\begin{equation}
G_{\mu\nu}+ \Lambda g_{\mu\nu}= T_{\mu\nu} \ ,
\end{equation} 
we perturb the metric around the dS background so that $g_{\mu\nu}=\bar{g}_{\mu\nu} + h_{\mu\nu}$  and define
\begin{equation}
\chi_{\mu\nu}=a^{-2}(\eta) \lp h_{\mu\nu} - \frac{1}{2} \bar{g}_{\mu \nu} h^\alpha_\alpha \rp 
\end{equation} 
where $a(\eta)$ is the scale factor. Now we choose the generalized harmonic gauge
\begin{equation}
\bar{\nabla}^\mu \chi_{\mu \nu}=0
\end{equation} 
in conformal coordinates where
\begin{equation}
\bar{g}_{\mu\nu} dx^\mu dx^\nu=a^2(\eta) (-d\eta^2+ d\rho^2 + \rho^2 d\Omega_2) \ 
\end{equation}
with $a(\eta)=-1/(H \eta)$ where $H$ is the Hubble's constant. Einstein equations become \cite{ashtekar15} 
\begin{align}
 \Box \bigg(\frac{\hat{\chi}}{\eta}\bigg) = - \frac{2 \hat{T}}{\eta} , \quad
\Box \bigg(\frac{{\chi_{0i}}}{\eta}\bigg)=- \frac{2 {T}_{0i}}{\eta}, \quad
\left(\Box + \frac{2}{\eta^2}\right) \left(\frac{\chi_{ij}}{\eta} \right)=-\frac{2}{\eta} T_{ij} \ .
\end{align}
Here $\hat{\chi}:=\chi_{00} + \chi_{ii}$, and similarly for $\hat{T}$. \\
\hfill \\ 
These can be solved using Green's functions. The most interesting case is the part with spatial indices. This can be solved as \cite{ashtekar15}
\begin{align}
\chi_{ij}(\eta, x) =&   \int d^3x' \frac{\eta}{|x - x'|(\eta-|x - x'|)} \left. T_{ij}(\eta', x')\right|_{\eta' = \eta - |x - x'|}+ \int d^3x' \int_{- \infty}^{\eta - |x
-x'|} d\eta' \frac{T_{ij}(\eta', x')}{{\eta'}^2} \ .
\end{align}
Now let us make some assumptions relevant for us to simplify these. We will be interested in results near $\mathcal{I}^+$, i.e. values of $\chi_{\mu \nu}(\eta,\rho,x^A)$ where $-\eta / \rho \ll 1 $. We will furthermore assume that we are far from the source: $x'<d \ll \rho$. Finally we will assume source to be slowly varying so that
\begin{equation}
T_{\mu\nu} \gg d \pr_\eta T_{\mu \nu} \gg \cdots \ .
\end{equation} 
Using these we will write for example
\begin{align}
\int d^3x' &\frac{1}{|x - x'| }  \left. T_{ij}(\eta', x') \right|_{\eta' = \eta - |x - x'|} = \int d^3 x'  \Bigg(  \frac{1}{\rho} \lp T_{ij}(\eta_r,x') + \vec n \cdot \vec{x}'\, T^{(1)}_{ij}(\eta_r,x')+\frac{1}{2} (\vec n \cdot \vec{x}')^2 T^{(2)}_{ij}(\eta_r,x')+\dots \rp \nonumber \\
&  +\frac{1}{\rho^2} \lp \vec n \cdot \vec{x}'\, T_{ij}(\eta_r,x')+\frac{3}{2} (\vec n \cdot \vec{x}')^2T^{(1)}_{ij}(\eta_r,x')+\dots \rp  +\frac{3}{2\rho^3} \lp (\vec n \cdot \vec{x}')^2 \, T_{ij}(\eta_r,x')+\dots \rp+O(\rho^{-4}) \Bigg)
\end{align}
where $\eta_r:=\eta-\rho$.

\subsection{Multipole Expansion and the Consistent Truncation}
\begin{wrapfigure}{r}{0.5\textwidth}
\centering
  \includegraphics[width=0.90\linewidth]{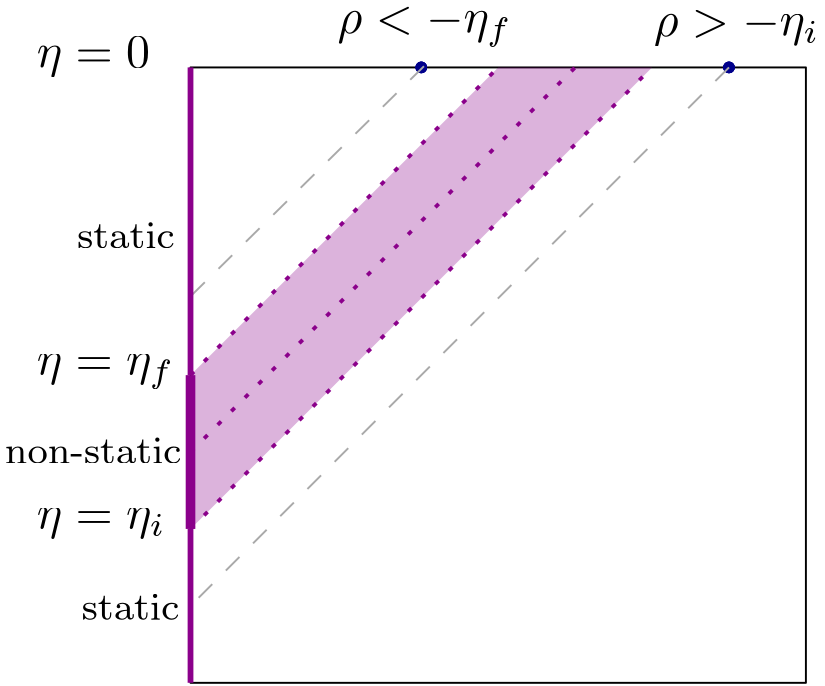}
  \caption{\label{penrose} {Penrose diagram of de Sitter with a source at the origin $\rho=0$, radiating for a finite duration.}}  
\end{wrapfigure}

Now we will express all of these contributions in terms of the multipole moments of the source.  We make the following definition
\begin{equation}
Q_L^{(\rho)}(\eta):= \int d^3x \ a^{\ell + 1}(\eta)\ T_{00} \ x_L \ ,
\end{equation}
for any positive integer $\ell$, where we define the multi-index $L=i_1 \cdots i_\ell$ and the notation $x_L=x_{i_1} \cdots x_{i_\ell}$.\\ 
Similarly we define
\begin{align} 
Q_L^{(p)}(\eta)&:= \int d^3x \ a^{\ell + 1}(\eta) \ T_{ii} \ x_L , \quad
P_{i \vert L}(\eta):= \int d^3x \ a^{\ell + 1}(\eta) \ T_{0i} \ x_L, \\
S_{ij \vert L}(\eta) &:= \int d^3x \ a^{\ell + 1}(\eta) \ T_{ij} \ x_L.
\end{align} 
Note that $Q_L^{(p)}(\eta)=S_{ii|L}$.  Note that not all of these are independent since the energy momentum conservation equations $\nabla^\mu T_{\mu\nu}=0$ yield
\begin{align}
    \dot Q_L^{(\rho)} &= H(\ell Q_L^{(\rho)}-Q_L^{(p)})- \ell  P_{(i_1|i_2 \cdots i_\ell)} \ ,\\ 
    \dot P_{i|L} &= (\ell-1) H P_{i|L} - \ell  S_{i(i_1|i_2 \cdots i_\ell)} \ .
\end{align}
Now we take another step and restrict our attention to contributions up to (and including) the quadrupolar order,  i.e. we will assume
\begin{equation}
\int d^3x \ a^{\ell+1} \ T_{\mu \nu} \ x_L = 0 \ \mbox{for} \ \forall\ell>2 \ ,
\end{equation} 
i.e. any moment with $L$ more than two indices is zero.  Using this in the energy momentum equations we will moreover have the conditions
\begin{align}
P_{(i \vert j k )}=0, \qquad P_{i \vert j k l}=0,\qquad S_{i (j \vert kl )}= 0, \qquad S_{(ij \vert k)}=0.
\end{align} 
which will further imply
\begin{align}
P_{i \vert jk} &=\frac{1}{2}\epsilon_{li(j}J_{k)l}-\frac{1}{2}\delta_{i(k}P_{j) \vert ll}+\frac{1}{2}\delta_{jk}P_{i \vert ll},\\
S_{i j \vert k} &= \frac{1}{2}\epsilon_{kl (i} K_{j)l}-\frac{1}{2}\delta_{k(i}Q^{(p)}_{j)}+\frac{1}{2}\delta_{ij}Q^{(p)}_k, \\
S_{i j \vert kl} &= \delta_{ij}Q^{(p)}_{kl}- (\delta_{i(k}Q^{(p)}_{l)j}+\delta_{j(k}Q^{(p)}_{l)i})+ Q^{(p)}_{ij}\delta_{kl}-\frac{1}{2} \delta_{ij}\delta_{kl}Q^{(p)}_{mm}+\frac{1}{2} \delta_{i(k}\delta_{l)j}Q^{(p)}_{mm},\label{Sijkl}
\end{align}
for some $J_{ij}$ and $K_{ij}$, called the odd parity moments, which are fixed uniquely by the decomposition. Thus we see the independent multipole moments we are left with are $Q^{(p)}, Q^{(p)}_i,Q^{(p)}_{ij},Q^{(\rho)},Q^{(\rho)}_i,Q^{(\rho)}_{ij}, J_{ij},K_{ij}$ and $P_{i|jj}$. In the literature people have kept $Q_{ij}^{(p)}\neq 0$ while letting $S_{ij|k\ell}=0$, but we see that this is inconsistent due to eq. \eqref{Sijkl}.\\
\hfill \\
Now we use these definitions in the expression for $\chi_{ij}$. We will focus on the time $\eta=0$. The result will be 
\begin{align} \label{eq:chiij}
\hspace{-0.8cm}\chi_{ij}(\eta=0,\rho,x^A) &=  \chi_{ij}(-\infty) +4H^{2} \bigg(\frac{1}{2} \dot Q_{ij}^{(\rho+p)}-HQ_{ij}^{(\rho)}+\frac{1}{2}n_{k}\epsilon_{kl(i}K_{j)l}-\frac{1}{2} n_{(i}Q_{j)}^{(p)}\nonumber \\ 
\hspace{-0.8cm} & \left. +\frac{1}{2} \delta_{ij} n_kQ_{k}^{(p)} +\frac{\delta_{ij}}{2} n_{k} n_{l} \dot Q_{kl} ^{(p)} - n_{l} n_{(i}\dot Q_{j)l}^{(p)} -\frac{1}{4} (\delta_{ij}-n_{i}n_{j}) \dot Q_{mm}^{(p)}
\bigg)  \right|_{\eta_r=-\rho}
\end{align}
where $n^i$ is the unit radial vector such that $\rho n^i=x^i$ and $\chi_{ij}(-\infty)$ is a constant. Now let us consider the case where our source is static except for a finite time $\eta_i>\eta>\eta_f$, see Fig.\ref{penrose}. That means for the times outside of this interval all the time derivatives of the multipole moments are zero. Then the gravitational radiation at time $\eta=0$ at the points $\rho_+ > -\eta_f$ or $\rho_- < -\eta_i$ can be written as
\begin{align}
\chi_{ij}^{TT}(\eta=0,\rho_\pm,x^A)=  \chi_{ij}(-\infty) +4H^{2} \left. \bigg( -HQ_{ij}^{(\rho)}+\frac{1}{2}n_{k}\epsilon_{kl(i}K_{j)l} \bigg) \right|_{\eta_r=-\rho_\pm }
\end{align}
Thus we see that there will be a difference in between the fields at the two extremities of positions $\rho=-\eta_f$ and $\rho=-\eta_i$: 
\begin{align}
\Delta \chi_{ij}^{TT}&= \chi_{ij}^{TT}(\eta=0,\rho=-\eta_f,x^A_f)-\chi_{ij}^{TT}(\eta=0,\rho=-\eta_i,x^A_i) \\ &= 4H^2  \bigg( -HQ_{ij}^{(\rho)}(\eta_f)+\frac{1}{2}n_{k}(x^A_f)\epsilon_{kl(i}K_{j)l}(\eta_f) + HQ_{ij}^{(\rho)}(\eta_i)-\frac{1}{2}n_{k}(x^A_i)\epsilon_{kl(i}K_{j)l}(\eta_i) \bigg) \ .
\end{align} 
One can show that this will produce a memory effect, namely the effect of the gravitational radiation remains even after the wave passes, see \cite{dS-rad-1} for more details. Note that this effect is $\mathcal{O}(H^2)$, however $\mathcal{O}(\rho^0)$ unlike the flat counterpart. The even parity part of $\chi_{ij}^{TT}$ at a single point can be gauged away, but not at both points, while the odd parity part cannot be gauged away even at a single point,  and is angle dependent \cite{dS-rad-1} .

\section{Charges}
\begin{wrapfigure}{r}{0.4\textwidth}
\centering
  \includegraphics[width=0.6\linewidth]{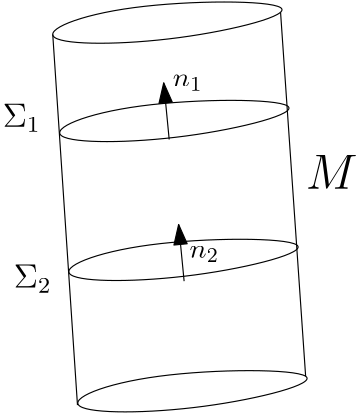}
  \caption{\label{man} A representative manifold on which we define a charge on the portion enclosed by the two boundaries.}
\end{wrapfigure}

Now we would like to get back to the question of charges. For our linearized construction above $E_{AB}$ is no longer independent of other variables, using this we will be able to define a proper charge.  
For the definition of charges we will use the following observation: Let $\mathcal{T}_{ab}$ be a tensor on a manifold $M$ with a metric $g_{ab}$ such that
\begin{equation}\label{T-rules}
g^{ab}\mathcal{T}_{ab}=0 \gag D^a \mathcal{T}_{ab}=0 \ .
\end{equation}
Let us also assume that there exist conformal Killing vector $\xi$ such that
\begin{equation}
D_{<a}\xi_{b>}=0 \ .
\end{equation} 
Then 
\begin{equation}
\int_{\Sigma_1} \epsilon_{\Sigma_1} n_a \mathcal{T}^{ab} \xi_b = \int_{\Sigma_2} \epsilon_{\Sigma_2} n_a \mathcal{T}^{ab} \xi_b
\end{equation}
for any $\Sigma_1,\Sigma_2$ that form a boundary of a manifold, see Fig.\ref{man}.
One can show that if we let
\begin{align}
   \mathcal{T}_{ab}=  \left[ \begin{array}{cc} -\frac{4}{3}M^{(\Lambda)} & -\frac{2}{3} N^{(\Lambda)}_B \nonumber \\ 
   -\frac{2}{3} N^{(\Lambda)}_B & J^{(\Lambda)}_{AB} + \frac{2}{\Lambda} M^{(\Lambda )} q_{AB} \end{array} \right] 
\end{align}
then \eqref{T-rules} is satisfied for 
\begin{equation}
g_{ab}dx^a dx^b= H^2 du^2+ q_{AB} dx^A dx^B \ ,
\end{equation}
the Bondi metric at $\scri^+$ by virtue of flux-balance laws. The tensor $\mathcal{T}_{ab}$ corresponds to the holographic stress-energy tensor on four-dimensional de Sitter \cite{Compere-LambdaBMS}. \\ 
For the case $q_{AB}=\mathring{q}_{AB}$ isometries of $dS_4$ are conformal Killing vectors at $\scri^+$, and the charges will be conserved in u. \\ 
We will use the same definition for the case $q_{AB} \neq \mathring{q}_{AB}$ to see how the charges flow.  We define
\begin{equation}
Q^T_\xi(u)=\int_{S^2_u} \epsilon_{S^2}  \mathcal{T}^{ub} \xi_b= \int_{S^2_u} \lp M^{(\Lambda)} \xi^u + \frac{1}{2} N_A^{(\Lambda)} \xi^A \rp
\end{equation}
Then we write down the $SO(1,4)$ charges by letting $\xi^a$ be the pullback of isometries to $\scri^+$.  We express them in terms of multipole moments by gauge transforming the metric perturbations in harmonic gauge, such as \eqref{eq:chiij}, to Bondi gauge. At linear order the Bondi variables are then given as
\begin{align}
\delta q_{AB} &= e^i_{\langle A} e^j_{B \rangle} \lp  \partial_u \zeta_{ij}+2H^2 \partial_u Q^{(\rho+p)}_{ij}+2H^2 n_k\epsilon_{kl(i}(K_{j)l}+H\int^u du'K_{j)l}(u')) \rp, \label{qABBondi}\\
\delta C_{AB} &= e^i_{\langle A} e^j_{B \rangle} \lp 3\zeta_{ij} +2 (\partial_u^2 -H^2)Q^{(\rho+p)}_{ij}+2n_k\epsilon_{kl(i} (\partial_u+H)K_{j)l} \rp ,\label{CABBondi} \\
\delta E_{AB} & = e^i_{\langle A} e^j_{B \rangle} \lp 2  Q^{(\rho+p)}_{ij} + 2n_k\epsilon_{kl(i} J_{j)l}. \rp \label{EABBondi}
\end{align}
where $e_A^i$ describes the projection from the cartesian coordinates $x^i$ to the angular coordinates $x^A$. Here $\zeta_{ij}(u)$ is defined as a solution to
\begin{align}
   \partial_u^2 \zeta_{ij}-3H^2 \zeta_{ij}=-2GH^{4}Q_{ij}^{(\rho + p)} \ , 
\end{align}
which pops up due to gauge transformation from the harmonic gauge to Bondi gauge. Using these in the definitions of $M^{(\Lambda)}$ and $N_A^{(\Lambda)}$ in \eqref{eq:M-Lambda} and keeping to quadratic order we will have the following results.
For dilatations $D=\partial/\partial u$, we have
\begin{equation}
Q^T_D= \int_{S^2_u} M^{(\Lambda)}
\end{equation}
and the flow can be shown to be \cite{dS-rad-2}
\begin{equation}
\dot{Q}^T_D = -{{\dQadd}}-\frac{G}{2}  \lp \dddot{Q}^2_{ij}+5H^2 \ddot{Q}^2_{ij}+4H^4 \dot{Q}^2_{ij} + \dddot J_{ij}^2+5H^{2} \ddot J_{ij}^2+4H^{4} \dot J_{ij}^2 \rp 
\end{equation} 
at second order in perturbations, where we identify an additional term $\Qadd_D$ in terms of the multipole moments,
\begin{equation}
\Qadd_D=  H^2 \ddot{Q}_{ij} \dot{Q}_{ij}+ H^4 \dot{Q}_{ij} {Q}_{ij} + \cdots + \ddot{J}_{ij} \dot{J}_{ij}+ \cdots
\end{equation} 
so that we get a positive mass loss with a redefinition of the charge: 
\begin{equation}\label{eq:Mass-loss}
\dot{Q}^F_D=-\frac{G}{2} \lp \dddot{Q}^2_{ij}+5H^2 \ddot{Q}^2_{ij}+4H^4 \dot{Q}^2_{ij} + \dddot J_{ij}^2+5H^{2} \ddot J_{ij}^2+4H^{4} \dot J_{ij}^2 \rp 
\end{equation}
where $Q^F_D:=Q^T_D+ \Qadd_D$.  So we see a modification of the naive proposal of $M^{(\Lambda)}$ as the Bondi mass aspect on dS.\\
Similar manipulations can be performed for rotations, boosts and translations.  For rotations 
\begin{align}
\dot{Q}^F_{R^i} = -\frac{2G}{5} \epsilon_{imn}  \bigg( & \ddot Q_{km}^{(\rho+p)}\dddot Q_{kn}^{(\rho+p)}+5H^{2} \dot Q_{km}^{(\rho+p)}\ddot Q_{kn}^{(\rho+p)}+4H^{4}  Q_{km}^{(\rho+p)}\dot Q_{kn}^{(\rho+p)} \nonumber \\ 
 & + \ddot J_{km}\dddot J_{kn}+5H^{2} \dot J_{km}\ddot J_{kn}+4H^{4}  J_{km}\dot J_{kn}\bigg),
\end{align} 
whereas for translations $\varepsilon=1$ and boosts $\varepsilon=-1$
\begin{equation}
\dot{Q}^F_{P^i,K^i}=-\frac{4G e^{\varepsilon Hu}}{{15}} \epsilon_{imt} (3 \varepsilon H^{3}  \zeta_{ml} \dot J_{lt}+2H^{6} Q_{ml} J_{lt}+ 6H^{4} \dot Q_{ml} \dot J_{lt}+6H^{2} \ddot Q_{ml} \ddot J_{lt}+ \dddot Q_{ml} \dddot J_{lt}) \ .
\end{equation} 
 We note that flat limit $H \ra 0$ of all of our charges matches the previous results found for the flat case e.g \cite{Bekenstein'73}, \cite{Thorne'74}. Moreover our de Sitter mass loss \eqref{eq:Mass-loss} matches that of Bonga et.al. \cite{Bonga-Bunster} calculated in a different gauge.

\section{Conclusions}
To summarize, in our papers \cite{dS-rad-1,dS-rad-2} with G.Compere and J.Hoque we have performed a consistent quadrupolar truncation of the multipole moments of a localized source on a dS background and written down metric perturbations near $\mathcal{I}^+$ for slowly varying sources. 
We have seen that a displacement memory effect appears by a passage of a gravitational wave, which is not subleading, differently than the flat case. Even parity contributions can be interpreted as a $\Lambda$-BMS vacuum transition while odd parity cannot be. We have defined charges that includes a proper mass, and matches the flat space formulae at the $H \ra 0$ limit.\\
We note however that the definitions of charges here including the mass are not unique as one can always add some total derivative to the left hand side. However the simplicity of the structure, especially for the cases of dilatation and rotations together with the matching the calculations in other gauge \cite{Bonga-Bunster} is encouraging. Also note that even though the generators of symmetries form a $SO(1,4)$ algebra, we have not shown that charges respects this algebra, this would be a work left for future.

%
%

\end{document}

%% file: globaldef.tex
\newcommand{\rp}{\right)}
\newcommand{\lp}{\left(}

\newcommand{\ra}{\rightarrow}

\newcommand{\pr}{\partial}

\newcommand{\gag}{\quad \mbox{and} \quad}